\documentclass[11pt]{article}
\usepackage{textcomp,latexsym,amssymb,amsmath,amsmath,graphicx}

\usepackage[numbers,sort&compress]{natbib}

\title{A Two-Dose Vaccine Epidemic Model with Power Incidence Rate}

\author{Gabriel O. Fosu\\ {\small Presbyterian University College, Ghana} \\
	{\small Email: gabriel.obed@presbyuniversity.edu.gh} \\
	 		 \vspace{1cm}
	 		 {\footnotesize Tel: +23320 2057 203} \\
	Emmanuel K. Mintah\\
	{\small Presbyterian University College, Ghana}}
\date{\empty}
\begin{document}
\maketitle

\begin{abstract}
	The dynamics of a SIVR model  with power relationship incidence rates ($ \beta I^{p} S^{q} $) is investigated. It is assumed an individual can be susceptible after receiving the first dose of the vaccine, hence a second dose is required to attain permanent immunity.  The steady states conditions of the  disease-free equilibrium and the endemic	equilibrium are critically  presented. Numerical simulations are carried out to determine the impact of the exponential parameters $ (p, q) $ on infection.\\


\end{abstract}

\section{Introduction}

A vast category of the conventional epidemiological models divides the host population $ (N) $  into few homogeneously mixed classes, usually susceptible (S), infected(I), exposed$ (E) $ and recovered$ (R) $. These models have the assumption that  the spread of  a disease is based on the principle of mass action transmission rate \cite{Anderson1991, Hethcote2000}. Anderson \& May \cite{Anderson1991} and  Hethcote \cite{ Hethcote2000} emphasized on a bilinear incidence rate between the susceptible class and infected class, $ \alpha S I $; where $ \alpha $ is the transmission coefficient. The model
$ \alpha S I $ assumes a completely homogeneous mix of the susceptible and infected population. 
An alternative bilinear incidence rate is the frequency-dependent transmission \cite{May1987},  that is  the susceptible class will acquire an infection based on the proportion
$ (I/N ) $. When the classes have a heterogeneous mix, there is the need to modify the standard incidence rate to a non-linear form. Other factors that necessitate a non-linear  transmission model are: high proportion of the infectious class \cite{Capasso1978,Brown1995}  and in a situation where there is a tendency of several infections of a host \cite{Korobeinikov2005}. 

In 1986, \cite{Liu1986} suggested a non-linear transmission of the form $ \dfrac{kI^{pS}}{1 + \alpha I^{l}} $.  Another non-linear form $ kS \ln (1 + v P/k) $ was proposed by \cite{Briggs1995} on the study of  insect–pathogen interactions in stage-structured populations.

Derrick \& van den Driessche \cite{Derrick2003}  and  Li \& Muldowney \cite{Li1995} explored the power incidence rate of the form $ \alpha I^{p} S^{q} $,  where $ 0<p\leq 1, \ 0< q\leq 1  $. Different values of $ p, q $ will result in varied situations. In this research work, the authors explored the model behavior of the SIVR model  with different parameter values for $ p, \mbox{ and } q $. Korobeinikov \& Maini \cite{Korobeinikov2005} did establish that a sufficient condition for global stability for a SIR model with this incidence rate is when $ p \leq 1 $. The parameter $ q $ does not affect the global properties of their system. Upon these, the researchers sought to identify the model behavior of a SIVR model under variable $ p $ and $ q $ values. A unique characteristic of this model under study is that, a proportion of the susceptible population have already been vaccinated.

\section{SIVR with Power Transmission}

The SIR epidemiological model was initially studied by Kermack and McKendrick \cite{Kermack1927}.
He divided the population into three different compartment susceptible (S), infected (I), and removed (R), with the assumption of constant population, that is total birth equals total death. With the SIVR model, $V$ represents the vaccinated individual density. These are individuals that have  gained temporal  immunity after the first dose.        
After the initial dosage, the individual moves back to the  susceptible class after the  failure of receiving the second dose of the vaccine. Permanent immunity in attained only after the second dose. 


Individuals receive the first dose at a rate $\psi$ which wanes at a rate $\omega$, the second dose is disbursed at the rate $\rho$. The birth or death rate is represented as $\mu$ which shows the magnitude of births or deaths in each compartment. The infection is carried on by the contact rate $\beta$. Also, individuals recover at a rate $\upsilon$.
The resulting differential equation is of the form:

\begin{equation}\label{eq1}
\begin{aligned}
\frac{dS}{dt} & = \mu N+\omega V-\beta I^{p}S^{q}-S(\psi+\mu)\\
\frac{dI}{dt} & = \beta I^{p}S^{q}-I(\mu +\upsilon)\\
\frac{dV}{dt} & = \psi S-\omega V-V(\mu+\rho)\\
\frac{dR}{dt} & = \upsilon I-\mu R+\rho V
\end{aligned}
\end{equation}


\section{Model Analysis}

Epidemiological models usually have two steady states; the disease free equilibrium and the endemic equilibrium. There are no cases of infection at the disease-free equilibrium, whilst, the equilibrium is endemic otherwise. However, compartmental models with power transmission rate can have either one/two steady states or no steady state \cite{Liu1987}.  The following is implemented to find the steady state solutions;
\[\frac{dS}{dt}= \frac{dI}{dt}=\frac{dV}{dt}=\frac{dR}{dt}=0\]
therefore equation \eqref{eq1} becomes;
\begin{equation}
\begin{aligned}
0 & = \mu N+\omega V-\beta I^{p}S^{q}-S(\psi+\mu)\\
0 & = \beta I^{p}S^{q}-I(\mu +\upsilon)\\
0 & = \psi S-\omega V-V(\mu+\rho)\\
0 & = \upsilon I-\mu R+\rho V
\end{aligned}
\end{equation}\label{eq3}

At the Disease Free Equilibrium (D.F.E.) there is no infection, hence no recovered people in the population(N).
We set $I^{o} = 0$ into the homogeneous set of differential equations to obtain;

\[ S^{o} = \dfrac{\mu (\omega + \mu + \rho)}{(\psi+ \mu)(\omega + \mu + \rho)- \psi \omega}, \quad V^{o} = \dfrac{ \psi \mu }{(\psi+ \mu)(\omega + \mu + \rho)- \psi \omega} \]

\[  \quad  R^{o} = \dfrac{ \rho  \psi }{(\psi+ \mu)(\omega + \mu + \rho)- \psi \omega}\]


as the disease-free equilibrium values. This is a typical setting where the susceptible population gets themselves vaccinated before an endemic eruption.

The Jacobian matrix at this state is;

\[J(S^{o},I^{o},V^{o},R^{o})=\left(\begin{array}{cccc}
-\mu-\psi & 0 & \omega  & 0 \\ 
0& -\mu-\upsilon & 0 & 0 \\ 
\psi & 0  & -\omega-\mu-\rho & 0 \\ 
0 & \upsilon & \rho & -\mu
\end{array}\right)\]

To determine the equilibrium we find roots of the characteristic equation $J-\lambda I=0$. The resultant eigenvalues are:

\begin{equation*}
\begin{aligned}
\lambda_{1} & =- \mu\\
\lambda_{2} & =-\mu- \upsilon\\
\lambda_{3} & =-(\psi+2\mu+\rho+\omega)+\sqrt{(\psi+2\mu+\rho+\omega)^{2}-((\mu+\psi)(\omega+\mu+\rho))}\\
\lambda_{4} & =-(\psi+2\mu+\rho+\omega)-\sqrt{(\psi+2\mu+\rho+\omega)^{2}-((\mu+\psi)(\omega+\mu+\rho))}
\end{aligned}
\end{equation*}

Since all the parameters are non negative,  all the above eigenvalues have negative real part; implying the disease free equilibrium is stable.

It is imperative to note that, if  the basic reproductive number $R_{0}$ is less than unity there exist a stable disease free equilibrium. Using the next generation matrix method, the reproductive number for this SIVR model is defined as: $R_{0}=\rho(\mathcal{KM}^{-1})$, where $ \mathcal{K} $ is the rate of appearance of a new infection and $ \mathcal{M} $ is the transfer rate of an infectious person in and out of a given cell \cite{Rizvi2016,VandenDriessche2002}.  Therefore,
$ \rho(\mathcal{KM}^{-1})=\rho\left(\frac{p\beta I^{p-1}S^{q}}{\mu+\upsilon}\right) $. With the assumption that $ p=q=1 $, then \[R_{0}=\dfrac{\beta S^{0}}{\mu+\upsilon} = \dfrac{\beta(\omega + \mu)}{(\mu +\upsilon)(\omega + \mu +\psi )} \]

At the endemic state, the Jacobian is defined as:

\begin{equation}
\begin{split}
&J(S^{*},I^{*},V^{*},R^{*})
= \\
&\left(\begin{array}{cccc}
-q\beta I^{*p}S^{*q-1}-\mu-\psi & -p\beta I^{*p-1}S^{*q} & \omega & 0 \\ 
q\beta I^{*p}S^{*q-1} & p\beta I^{*p-1}S^{*q}-\mu-\upsilon & 0 & 0 \\ 
\psi & 0 & -\omega-\mu-\rho & 0 \\ 
0 & \upsilon & \rho & -\mu
\end{array}\right)
\end{split}
\end{equation}

The endemic equilibrium state $ E^{*} $ is such that 

$ S^{*}({\psi+\mu})  = {\mu +\omega V -I^{*}(\mu + \upsilon)}, $\quad
$ I^{*}({\mu+\upsilon})  ={\beta I^{*p}S^{*q}} $, \quad
$ V^{*} = \dfrac{\psi S^{*}}{\omega+\mu+\rho} $, \quad
$ R^{*}  = \dfrac{\upsilon I^{*}+\rho V^{*}}{\mu} $

The characteristic equation for the above Jacobian matrix is deduced to: $ \lambda_{1} = - \mu $ and
\begin{equation}
\lambda^{3}-  \lambda^{2}(\kappa+ \tau + \delta)+ \lambda ( \kappa \tau + \kappa \delta + \tau \delta + mn-\omega\psi)-\kappa \tau \delta- mn\delta+ \omega \psi \tau=0 \label{e2}
\end{equation}

where $ m =q\beta I^{*p}S^{*q-1}, \ n{qI^{*}} ={mpS^{*}}, \ 
-\kappa = \mu + m + \psi,  \ \tau = n - \mu - \upsilon, \ \mbox{and } -\delta = \omega + \mu + \rho   $

From numerical computation,  we realized the real part of equation \ref{e2} can not be positive. This indicates that, the steady state(s) will also be stable.  

\section{Numerical Simulation}
The graphical outputs were produced with the following assumed compartmental values: $ N = 10,000 $, with 40\% as total susceptible population; 30\% infected, 15\% vaccinated and 15\% in the recovered class. 

\begin{figure}[hp]
	\centering
	\includegraphics[height=6cm, width=13cm]{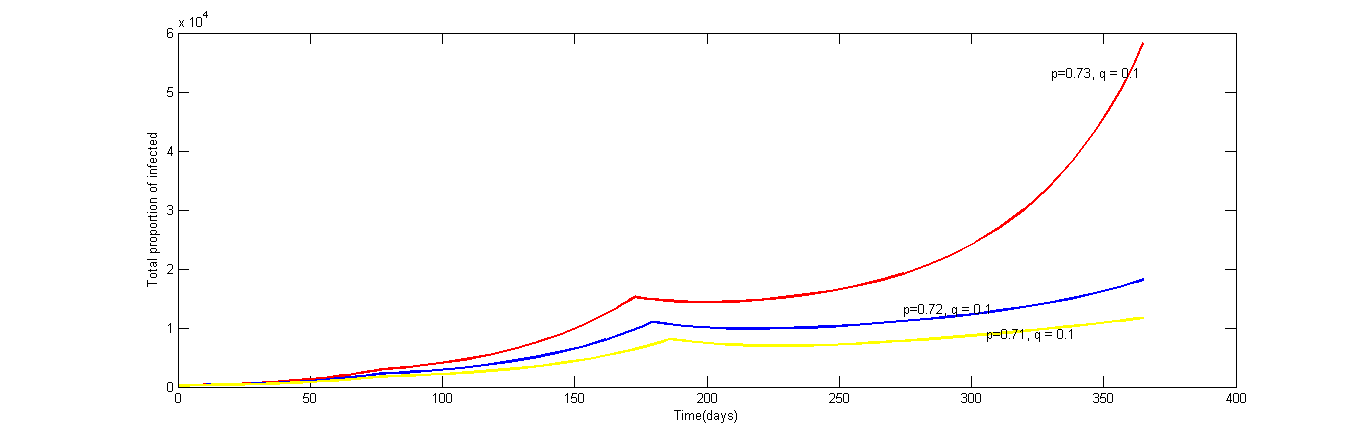}
	\caption{Results for $ p > 0.7  $ with $ q=0.1$}
	\label{f2}
\end{figure}

\begin{figure}[hp]
	\centering
	\includegraphics[height=6cm, width=13cm]{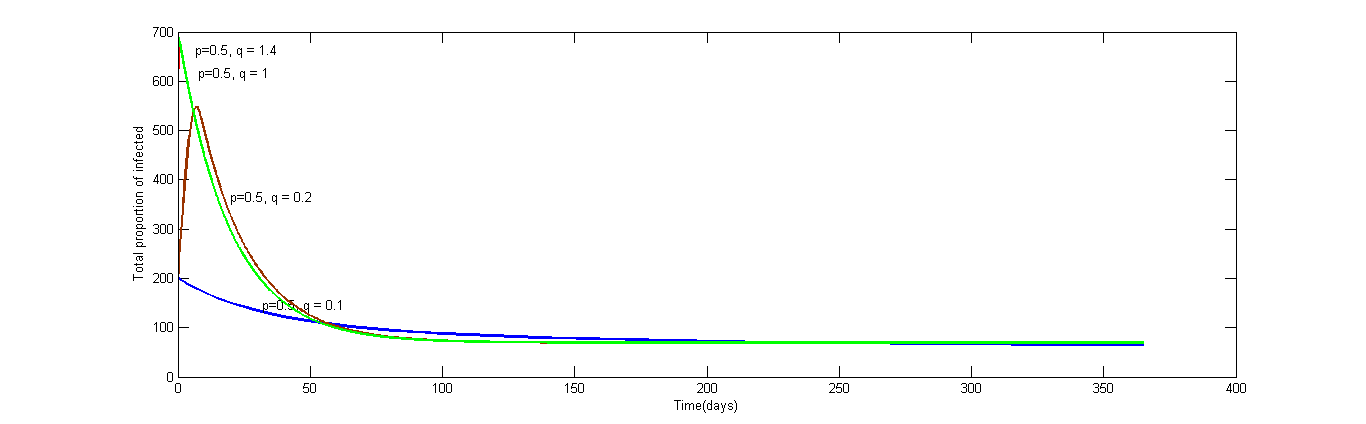}
	\caption{Results for $ p< 0.7 $ with a variant of $ q $}
	\label{f3}
\end{figure}

\begin{figure}[hp]
	\centering
	\includegraphics[height=6cm, width=13cm]{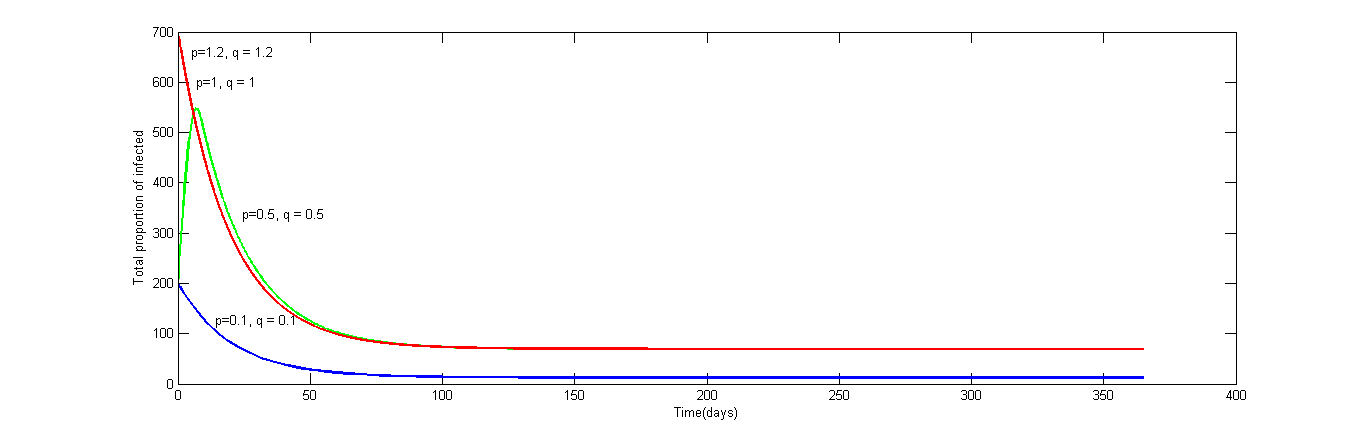}
	\caption{Graphical result when $ p = q $}
	\label{f4}
\end{figure}

Satisfactory parameter values indicating the presence of the endemic are $ \mu= 0.0035343,\ \psi = 0.00020185263, \ \upsilon = 0.0476, \ \beta = 0.26199 , \ \rho= 0.0054795, \ \omega = 0.0027  $ \cite{Rizvi2016}. 

With the graphical simulation, we realized that when $ p > 0.7 $, the infectious population remains boundless as $ q \rightarrow 0 $, see figure \ref{f2}. Whereas the simulation for $ p<0.7 $ is shown in figure \ref{f3}. $ p $ is chosen as 0.5 along with a variant of $ q $ values.  In figure \ref{f4} the values of $ p $ and $ q $ are equal to each other. When both values are closer to zero the model has a different equilibrium vis-a-vis values closer to one or  greater than one.

\section{Conclusion}
As a result of several endemic diseases being resistant to its drug, the SIVR model with nonlinear incidence rate is analyzed; hypothesizing a multiple vaccine case. The model considered  a double dose vaccine  for an individual to attain permanent immunity. The model had two equilibrium states; the disease free equilibrium and the endemic equilibrium. These states were all found to be stable. At the endemic state, the infected class converges to  a fixed  number. However, when the exponential $ p $ and $ q $  are close to zero the infection converges quickly. The infection grows unboundedly when $ p $ is greater than 0.7 with $ q $ close to zero. 

\bibliography{sample}
\bibliographystyle{plannat}

\end{document}